\documentclass[12pt]{book}

\usepackage[dvips]{graphicx,color}
\usepackage{makeidx,tsukuba}

\makeauthorindex
\makeindex

\begin{document}

\BookTitle{\itshape The 28th International Cosmic Ray Conference}
\CopyRight{\copyright 2003 by Universal Academy Press, Inc.}
\pagenumbering{arabic}

\chapter{
Superluminal Particles, Cosmology \\
and Cosmic-Ray Physics}

\author{%
%
%
Luis Gonzalez-Mestres\\
{\it LAPP, CNRS-IN2P3, B.P. 110 , 74941 Annecy-le-Vieux Cedex}\\
}

\section*{Abstract}
Non-tachyonic superluminal sectors of matter (superbradyons), with critical 
speeds in vacuum much larger than the speed of light, can quite naturally 
exist and play an important role in both cosmic-ray physics (anomalous 
high-energy events) and cosmology (big-bang physics, alternatives to 
inflation, dark matter...). They can even be the real "elementary" 
particles. An updated discussion of the subject is 
presented, in relation with recent theoretical and experimental results. 
Prospects for future searches are also reexamined. Lorentz symmetry violation
(LSV) models based on mixing with superbradyons are compared with LDRK (linearly 
deformed relativistic kinematics) and QDRK (quadratically deformed relativistic 
kinematics) such as defined in our previous paper physics/0003080 .  

\section{Superbradyons}

Introduced in [3], superbradyons are superluminal particles with positive 
mass and energy, 
and critical speeds in vacuum much larger than the speed
of light $c$ (see [3-13] and references therein). They can be assumed 
to be the actual building blocks of matter and to provide an alternative
to inflationary cosmological models. They can 
also be invoked to explain the absence of
the Greisen-Zatsepin-Kuzmin (GZK) 
cutoff or be the basic ingredient of new models to solve the dark matter
problem. Present low-energy bounds on LSV do not
allow to exclude the possible existence of such particles 
and, to date, no fundamental
argument has been provided to preclude their existence.
We therefore must consider the possibility that superbradyons
exist and play a real 
dynamical and cosmological role. The energy $E$ and momentum $p$
of a superbradyon with mass $m$ and critical speed $c_i~\gg ~c$
will be given by the generalized relativistic equations:
\begin{eqnarray}
p~=~m~v~(1~-~v^2c_i^{-2})^{-1/2}~ \\
~E~=~m~c_i^2~(1~-~v^2c_i^{-2})^{-1/2} \\
E_{rest}~~=~~m~c_i^2~~~~~~~~~~~~~~~
\end{eqnarray}
\noindent
where $i$ stands fot the $i$-th superluminal sector of matter, $v$ 
is the speed and $E_{rest}$ the rest energy.
Each superluminal sector of matter will have its own Lorentz invariance
with $c_i$ defining the metric.
We call the sector made of particles with critical speed in
vacuum = $c$ the "ordinary" sector (for "ordinary" particles
or "bradyons"). Being able to write 
formulae (1-3) for all sectors simultaneoulsy requires the
existence of an "absolute" rest
frame ({\bf the vacuum rest frame}, VRF, possibly close to that
suggested by the study of cosmic microwave background radiation),
the only one where this will be
possible.
Furthermore, interactions between two different
sectors of matter will break the two sectorial 
Lorentz invariances and deform their
associated relativistic kinematics.

\section{New Phenomenological Issues}

A specific property of LSV induced in the "ordinary" sector of matter
by the mixing with 
superbradyons is that the LSV energy dependence can follow many 
different patterns
according to the details of the mixing. 
It may even happen that the effective LSV parameter (basicallly, the deformation term in
the hamiltonian divided by the squared momentum scale) induced by the mixing
decreases after some
critical energy instead of growing [11] as in LDRK (linearly with energy) 
and QDRK (quadratically with energy).  
There are several recent phenomenological controversies where superluminal particles 
are potentially able to play a decisive role (see [11-13] and references 
therein): 

\subsection{Cosmological Issues}

From a theoretical
point of wiew, as cosmological 
observations appear to confirm some basic aspects of the inflationary 
scenario
(see [16] and references therein), it seems compelling that any alternative 
to standard inflation
incorporates two essential properties: a) superluminal expansion; b) entropy 
production.
Big-bang models where "ordinary" matter would be composite and made of superluminal
particles present interesting potentialities in the field. It is even 
possible [6,12,13] to postulate the existence of a fundamental length scale
$a_i$ for each superluminal sector of matter, besides the usual (Planck) scale 
$a$ associated to the "ordinary" sector. We expect $a_i~\ll ~a$ .
Our Universe may then originate from 
a set of transitions at the $a$ and $a_i$ scales where "ordinary"
matter cannot exist at length scales below $a$ , and similarly for the $i$-th sector 
of matter below $a_i$ . 

\subsection{The TeV Region}

Recent speculations from extragalactic astronomy
that photons in the TeV region may be being observed when, according to special 
relativity, they 
should not, have led to interpretations based on LSV models   
of the LDRK type (linearly deformed relativistic kinematics, see [11]). 
But these models (see [2,14,15] and references therein), where the LSV 
parameter increases linearly with energy, lead 
to severe consistency problems related to the global implications 
of LDRK [11]. Instead, there may be at least two suitable
alternatives to LDRK
based on superluminal particles [12,13]:

a) Superluminal particles are emitted by gamma factories and emit in turn "Cherenkov"
radiation ("ordinary" or superluminal particles) or decay later into 
photons which reach the detector.

b) There is a very small mixing between the photon and a superluminal particle, 
vanishing in the zero-momentum limit, reaching its maximum at TeV 
energies and then decreasing for some dynamical reason. 
Such a mixing would lead to LSV effects in the photon propagator at TeV energies, but  
the effective LSV parameter would not rise as energy increases further and 
global inconsistencies can in principle be avoided. 

More generally, suggestions that LSV may produce observable effects 
at energies far below those of the 
highest-energy cosmic rays seem difficult to reconcile in a consistent way
with standard deformed relativistic kinematics (DRK) models where the parameter 
driving LSV would vary like a power of momentum. But an effect due to mixing with 
the superluminal sectors of matter, leading to a very small LSV with observable 
effects for a given "ordinary" 
particle in a precise energy range, can be an alternative to these models.  
 
\subsection{The UHECR Region}

A similar phenomenological pattern can be imagined for events originating 
from ultra-high energy cosmic rays (UHECR). 
From an experimental point of view,
it is not clear by now whether the GZK 
cutoff exists or not. It is even not excluded that the cutoff be delayed (see [1] and 
references therein), so that it would not be at work below an energy threshold higher 
than that expected from standard theoretical calculations. 

If the GZK mechanism is just delayed, a very small mixing 
of "ordinary" UHECR with 
superluminal particles [12,13] can
account for such a behaviour similarly to the previous sub-section, and reproduce 
the delayed cutoff. An important conceptual question is how the internal structure
of the "ordinary" ultra-high energy particle (UHEP) would be altered by such a mixing. 

\subsection{Direct Signatures}  

As discussed in [4,8,10], the direct signature of a superbradyon with $c_i ~\gg ~c$ 
colliding with any terrestrial  
absorber, including the atmosphere, would be unique because of the kinematics (1-3). 
The main feature of the event would be its lack of directionality, due to the very large
incoming $E ~/~p$ ratio. The only possible background would be a non-relativistic 
"ordinary" neutral superheavy object, but such an exceptional competitor can also be ruled
out in a certain number of cases. Superluminality would often be impossible to fake, 
especially if the superbradyon is relativistic 
with respect to its own superluminal sector and if there are several events [12,13]. 
Cherenkov radiation in vacuum, interactions or decays of superluminal secondaries,
energy spectra of the observed events... would be crucial for event analysis.  

\subsection{Conclusion and Comments}

LSV models based on a small mixing with superbradyons may provide original 
solutions to present phenomenological controversies and puzzles concerning 
not only basic cosmology or dark matter, but also
the properties of high-energy and ultra-high energy cosmic rays. Thus, rare 
and anomalous vents in UHECR cosmic-ray experiments or TeV gamma physics may
potentially provide clues to the understanding of fundamental Big-Bang dynamics.

It therefore
seems also necessary to 
complete previous studies ([4,6,10,11] and references therein) and 
further explore the implications of superbradyons for Planck-scale dynamics, string
models and other possible ingredients of current cosmological models. 
The validity of standard quantum mechanics for superluminal sectors  
is not obvious, although we have always assumed. In particular, 
the question of the universality of the Planck constant is worth exploring.  

More details on the ideas presented here can be found in references [12] and [13],
as well as in subsequent
papers of the same series ({\it Deformed Lorentz symmetry and High-Energy
Astrophysics}, see arXiv.org ).

\section{References}
\re
1.\ Amelino-Camelia G.\ 2002a, paper gr-qc/0209232 of arXiv.org
\re
2.\ Amelino-Camelia G.\ 2002b, paper gr-qc/0212002 of arXiv.org
\re
3.\ Gonzalez-Mestres L.\ 1995, paper hep-ph/9505117
of arXiv.org
\re
4.\ Gonzalez-Mestres L.\ 1996, paper hep-ph/9610474 of arXiv.org
\re
5.\ Gonzalez-Mestres L.\ 1997a, paper physics/9702026 of arXiv.org
\re
6.\ Gonzalez-Mestres L.\ 1997b, paper physics/9704017 of arXiv.org
\re
7.\ Gonzalez-Mestres L.\ 1997c, paper physics/9705032 of arXiv.org
\re
8.\ Gonzalez-Mestres L.\ 1997d, paper physics/97121049 of arXiv.org
\re
9.\ Gonzalez-Mestres L.\ 1997e, paper physics/9712056 of arXiv.org
\re
10.\ Gonzalez-Mestres L.\ 1999, paper hep-ph/9905454 of arXiv.org
\re
11.\ Gonzalez-Mestres L.\ 2000, paper physics/0003080 of arXiv.org
\re
12.\ Gonzalez-Mestres L.\ 2002a, paper hep-th/0802064 of arXiv.org
\re
13.\ Gonzalez-Mestres L.\ 2002b, paper hep-th/0210141 of arXiv.org
\re
14.\ Jacobson T., Liberati, S., Mattingly, D.\ 2002, paper astro-ph/0212190 of arXiv.org
\re
15.\ Jacobson T., Liberati, S., Mattingly, D.\ 2003, paper astro-ph/0303001 of arXiv.org 
\re
16.\ Turner M.S.\ 2002, paper astro-ph/0212281 of arXiv.org
\endofpaper
\end{document}